\documentclass{elsart}

\usepackage{amsmath}
\usepackage{amssymb}
\usepackage{graphicx}

\newcommand{\rmC}{\mathrm{C}}
\newcommand{\rmT}{\mathrm{T}}
\newcommand{\rmLS}{\mathrm{LS}}
\newcommand{\rmCoul}{\mathrm{Coul}}
\newcommand{\rmTE}{\mathrm{TE}}
\newcommand{\rmE}{\mathrm{E}}
\newcommand{\rmO}{\mathrm{O}}
\newcommand{\rmm}{\mathrm{m}}
\newcommand{\rmc}{\mathrm{c}}
\newcommand{\rmps}{\mathrm{p}}
\newcommand{\rmP}{\mathrm{P}}
\newcommand{\rmV}{\mathrm{V}}

\begin{document}
\runauthor{Sugimoto, Ikeda and Toki}
\begin{frontmatter}
\title{Charge- and parity-projected Hartree-Fock method for the strong tensor
correlation and its application to the alpha particle}
\author[RIKEN]{Satoru Sugimoto}
\ead{satoru@riken.jp}
\author[RIKEN]{Kiyomi Ikeda}
\ead{k-ikeda@riken.jp}
\author[RCNP]{Hiroshi Toki}
\ead{toki@rcnp.osaka-u.ac.jp}
\address[RIKEN]{Institute of Physical and Chemical Research (RIKEN),\\
Wako, Saitama 351-0198, Japan}
\address[RCNP]{Research Center for Nuclear Physics (RCNP), Osaka University,\\
Ibaraki, Osaka 567-0047, Japan}
\begin{abstract}
We propose a new mean-field-type framework which can treat the
strong correlation induced by the tensor force. To treat the
tensor correlation we break the charge and parity symmetries of a
single-particle state and restore these symmetries of the total
system by the projection method. We perform the charge and parity
projections before variation and obtain a Hartree-Fock-like
equation, which is solved self-consistently. We apply the
Hartree-Fock-like equation to the alpha particle and find that by
breaking the parity and charge symmetries, the correlation induced
by the tensor force is obtained in the projected mean-field
framework. We emphasize that the projection before the variation
is important to pick up the tensor correlation in the present
framework.
\end{abstract}
\begin{keyword}
tensor force \sep parity mixing \sep charge mixing \sep
Hartree-Fock approximation \sep variation after projection \sep
alpha particle
\PACS 21.10.Dr \sep 21.10.Gv \sep 21.30.Fe \sep 21.60.Jz
\end{keyword}
\end{frontmatter}

\section{Introduction}
%%%%%%%%%%%%%%%%%%%%%%%%%%%%%%%%%%%%%%%%%%%%%%%%%%%%%%%%%%%%%%%%%%%%%%%%
Mean-field models with effective interactions like the Skyrme
force \cite{vautherin72} or the Gogny force \cite{decharge80}, are
widely applied to various fields in nuclear physics with great
success. The relativistic mean field theory is also applied to
many nuclear systems successfully \cite{walecka74,sugahara94}.
These models are based on the single-particle picture of the
nuclear many-body system. In this picture, a nucleon in a nucleus
is moving in a nuclear mean field made by other nucleons in the
nucleus. The single-particle picture is known to be appropriate in the wide range
of nuclear phenomena since the establishment of the shell model
\cite{mayer49}.

Usually, the effective interactions used in nuclear mean-field
models only have the central and LS parts as the nuclear two-body
interactions. They do not have the tensor part. However, the
tensor force is known to be very important in the structure of
nuclei
\cite{akaishi72,akaishi86,carlson98,pieper01,suzuki98,terasawa60,bethe71}.
The tensor force is mainly mediated by the pion, which is one of
the most important meson since Yukawa proposed it as a mediator of
the nuclear force \cite{yukawa35}. Many studies have demonstrated
that the tensor force gives large attractive energy in light
nuclei \cite{akaishi72,akaishi86,carlson98,pieper01,suzuki98}. The
shell model calculation shows that the tensor force can produce
about a half of single-particle \textit{ls} splittings in light
nuclei \cite{terasawa60}. These facts show that the tensor force
is an essential ingredient of the structure of light nuclei.
Furthermore, the tensor force plays an important role in the
saturation mechanism of nuclear matter \cite{bethe71}. This fact
indicates that the tensor force may have an important effect also
on the structure of nuclei in the heavy mass region.

The importance of the tensor force mentioned above inspires us to
treat the tensor force seriously even in a mean-field model. Many
people, however, believe that the effect of the tensor force can
be treated by renormalizing the central and LS parts of the
effective two-body interaction.  This renormalization is founded by
the G-matrix theory, where the effective interaction is made by
renormalizing the central and LS forces to include the effect of
the tensor force together with the short-range correlation. In the
G-matrix theory, the effective tensor force is weakened and
treated simply as the residual interaction. The main
part of the tensor force is included in the $^3$E part of the
central force. We see the importance of the tensor force
by comparing the theoretical results for nuclear matter and the alpha
particle. In nuclear matter, the ratio of the energy
contribution from the $^3$E part of the effective interaction and
that from $^1$E is nearly equal to 1, but in the alpha particle,
this ratio increases to about 1.5 \cite{akaishi72}. This
enhancement of the energy contribution from the $^3$E component
indicates that the effect of the tensor force is larger in the
alpha particle than in nuclear matter, and the tensor force
should be more important in finite nuclei than in nuclear
matter.

There are some Hartree-Fock calculations using the nuclear forces
which include the tensor force \cite{wong68,tarbutton68,rouben72}
or the pion \cite{brockmann78,bouyssy87} explicitly. They show
that the tensor force gives a large contribution to the
\textit{ls} splitting of single-particle states especially in the
spin-unsaturated nuclei. In contrast to the conventional
mean-field calculations, we recently proposed a new method to
treat the pion in the relativistic mean field (RMF) theory
\cite{toki02}. Because the pion is a pseudoscalar meson, a
single-particle state of a nucleon in a nuclear mean-field should
change its parity when the nucleon emits or absorbs a pion. It
suggests that we can incorporate the pion into the RMF theory by
introducing parity-mixed single-particle states. The idea of the
parity-mixed single-particle state was already discussed in the
references \cite{amiet63,amiet66,rohl66,bleuler66,bassichis67}
many years ago. In the previous paper \cite{toki02}, we applied
this idea in the RMF theory and found that the pion mean field
becomes finite, due to which the correlation induced by the pion
is appropriately taken into account. This result is not restricted
to a relativistic framework. We can treat the tensor force by the
introduction of the parity-mixed single-particle states in a
non-relativistic framework also.

The purpose of this paper is to apply the same idea in a
non-relativistic mean-field model and see the effect of the tensor
force in the alpha particle. We make two major improvements
compared to the previous study in the treatment of a nuclear
many-body system. The first improvement is straightforward as
coming from the isovector character of the pion. Because the pion
is a isovector meson, a single-particle state in a mean field
changes its charge state when it emits or absorbs a charged pion.
This fact indicates that the isovector nature of the pion can be
treated by introducing charge-mixed single-particle states, which
have both proton and neutron components. The dominant part of the
tensor force is $\boldsymbol{\tau}_1\cdot\boldsymbol{\tau}_2$-type
and therefore this improvement is important when we treat the
tensor force. The second improvement is related with the
restoration of symmetry. We mix the parities and the charges into
a single-particle state. A total wave function made from the
single-particle states with the parity and charge mixings does not
have a good parity and a definite charge number. Therefore we need
to perform the parity projection and the charge (number)
projection to obtain a total wave function having a good parity
and a definite charge number. By performing the projections we can
pick up the 2-particle--2-hole, the 4-particle--4-hole, $\dots$
correlations, which are induced by the tensor force for the ground
state of even-even mass nuclei \cite{toki02}, and are important in
the binding mechanism from light nuclei to nuclear matter.

We may think of performing the charge and parity projections after
we obtain a total wave function assuming the charge and parity
mixings (projection after variation). Because the effect of the
tensor force is strong and affect the binding mechanism of nuclei
largely, we need to treat the restoration of the parity and charge
symmetries carefully. Therefore, in a more sophisticated way, we
should first perform the parity and charge projections, and then
take a variation of the energy expectation value which is
evaluated with the projected total wave function (variation after
projection (VAP)). In the present study we perform calculations
for the alpha particle in the VAP scheme and show that the VAP
scheme is needed to treat the tensor force more appropriately. We
should note that the present framework is not a simple mean-field
model because we effectively superpose many Slater determinants by
performing the projections. In this sense, our framework is a kind
of descriptions beyond the mean field
\cite{takami96,valor00,rodriguez-guzman00,rodriguez-guzman02,bender03}.

This paper is arranged as follows. In Section~\ref{sec: cpphf}, we will formulate the charge- and
parity-projected Hartree-Fock method. In Section~\ref{sec:
result} we will apply the charge- and parity-projected
Hartree-Fock method to the alpha particle to show the
effectiveness of our method. In the last section we will give
the summary.

\section{Charge- and parity-projected Hartree-Fock
method}\label{sec: cpphf} We present the charge- and
parity-projected Hartree-Fock (CPPHF) method. In the CPPHF method,
we construct a total wave function with single-particle states
with the parity and charge mixings. A single-particle wave
function with the charge and parity mixings in the spherical case
is written as,
\begin{eqnarray}
 {\psi_{njm}}(x) =
  \sum_{t_z=\pm 1/2}\left(
             {
  \phi_{n j l t_z}(r) \mathcal{Y}_{j l m}(\Omega)\zeta(t_z)}
  +{
  \phi_{n j \bar{l} t_z}(r) \mathcal{Y}_{j \bar{l} m}(\Omega)\zeta(t_z)}
           \right)
  .
  \label{eq: spwf}
\end{eqnarray}
This wave function consists of four terms, each of which has a
positive or negative parity, and a charge number, 1 or 0 (proton
or neutron). Here, the radial wave functions $\phi$'s depend only
on the radial coordinate $r$, the isospin wave functions are
denoted as $\zeta(t_z)$ ($t_z=1/2$ for proton and $t_z=-1/2$ for
neutron), and $\mathcal{Y}_{jlm}$ is the eigenfunction of total spin
$\boldsymbol{j}=\boldsymbol{l}+\boldsymbol{s}$.
In the spherical case with the parity and charge mixings, the good
quantum numbers are $j$ and $m$. Two angular wave functions
${\mathcal{Y}}_{j l m}$ and  $\mathcal{Y}_{j \bar{l} m}$ are
related to each other as
\begin{eqnarray}
 {\mathcal{Y}_{j\bar{l} m}}(\Omega)
  =
  {\boldsymbol{\sigma} \cdot \Hat{\boldsymbol{r}}}{\mathcal{Y}_{j l m}}(\Omega),
  \qquad
  \bar{l} = \begin{cases}
              l + 1 & (j = l + \frac{1}{2})\\
              l -1 &  (j = l - \frac{1}{2})
             \end{cases}.
\end{eqnarray}
Here, $\boldsymbol{\sigma}$ is the Pauli spin operator and
$\hat{\boldsymbol{r}}$ is the unit radial vector. Because
$\boldsymbol{\sigma} \cdot \Hat{\boldsymbol{r}}$ is a $0^-$
operator, ${\mathcal{Y}}_{j l m}$ and ${\mathcal{Y}}_{j \bar{l}
m}$ have the same total angular momentum $j$ but different orbital
angular momenta, \textit{i.e.}, $|l-\bar{l}|=1$. It means that
they have the opposite parities to each other. The symbol $n$ is
introduced to label the single-particle states with the same $j$
and $m$. In the CPPHF method, $n$ does not correspond to the node
quantum number because of the parity and charge mixings. We take a
Slater determinant made from the charge- and parity-mixed
single-particle states as an intrinsic wave function for a nucleus
with the mass number $A$:
\begin{eqnarray}
 \Psi^{\textrm{intr}}=\frac{1}{\sqrt{A !}}
  \hat{\mathcal{A}} \prod_{a=1}^{A}&
  \psi_{\alpha_a} (x_a).
  \label{eq: wf intr}
\end{eqnarray}
Here, $\alpha_a$ denotes $n$, $j$, and $m$ in (\ref{eq: spwf}) and
$\hat{\mathcal{A}}$ is the antisymmetrization operator. As already
mentioned above, $\Psi^{\textrm{intr}}$ does not have a good
parity and a definite charge number. We perform the projections of
parity ($\pm$) and charge number ($Z$) on $\Psi^{\textrm{intr}}$;
\begin{align}
 \Psi^{(\pm;Z)} = \hat{\mathcal{P}}^\mathrm{p}(\pm)
 \hat{\mathcal{P}}^\mathrm{c}(Z) \Psi^{\textrm{intr}}
 .
\end{align}
Here, $\hat{\mathcal{P}}^\mathrm{p}(\pm)$ is the parity-projection
operator, where $\hat{\mathcal{P}}^\mathrm{p}(+)$ projects out the
positive parity state and $\hat{\mathcal{P}}^\mathrm{p}(-)$
projects out the negative parity one.
$\hat{\mathcal{P}}^\mathrm{c}(Z)$ is the charge-number-projection
operator, which projects out the wave function with a charge
number $Z$. Therefore, $\Psi^{(\pm;Z)}$ has a good parity ($\pm$)
and a definite charge number ($Z$). The parity projection operator
$\hat{\mathcal{P}}^\mathrm{p}(\pm)$ is defined as
\begin{align}
 \hat{\mathcal{P}}^\mathrm{p}(\pm) = \frac{1\pm \hat{P}}{2}
 \quad
  \left(\hat{P}=\prod_{a=1}^A \hat{p}_a
  \right)
  ,
 \label{eq: parityop}
\end{align}
where the total parity operator $\hat{P}$ is the product of the
parity operator  $\hat{p}_a$ for each single-particle state.
The charge projection operator $\hat{\mathcal{P}}^\mathrm{c}(Z)$
is defined as
 \begin{align}
 \hat{\mathcal{P}}^\mathrm{c}(Z) = \frac{1}{2\pi} \int_0^{2\pi} d \theta
  e^{i(\hat{Z}-Z)\theta}
  = \frac{1}{2\pi} \int_0^{2\pi} d \theta
  e^{-i Z \theta}\hat{C}(\theta)
  \quad
  \left(
 \hat{Z} = \sum_{a=1}^A \frac{1+\tau^3_a}{2}
 \label{eq: chargeop}
    \right)
    ,
 \end{align}
where $\hat{Z}$ is the charge number operator, which is the sum of
the single-particle proton projection operator $(1+\tau_a^3)/2$,
and the charge-phase operator is defined as
$\hat{C}(\theta)=e^{i\hat{Z}\theta}$.

We take a Hamiltonian $\hat{H}$ in the following form,
\begin{eqnarray}
 \hat{H} &=& - \sum_{a=1}^A \frac{\hbar^2}{2M}\triangle_a
  - \frac{1}{2 A M}\left\{\sum_{a=1}^A{(\frac{\hbar}{i} \boldsymbol{\nabla}_a)}\right\}^2
  \nonumber \\
  &&+\sum_{a > b = 1}^A(\hat{v}_\rmC(x_{ab})
   +\hat{v}_\rmT(x_{ab})+\hat{v}_{\rmLS}(x_{ab})
  +\hat{v}_{\rmCoul}(x_{ab}))
  \label{eq: hamiltonian1}\\
 &=& \sum_{a=1}^A \hat{t}(x_a)
  +\sum_{a > b = 1}^A \hat{v}(x_{ab})
  \label{eq: hamiltonian2}
  .
\end{eqnarray}
Here, the first and the second terms on the right-hand side are
the single-particle kinetic energy and the energy of the center of
mass motion. The two-body interactions, $\hat{v}_\rmC$,
$\hat{v}_\rmT$, $\hat{v}_\rmLS$, and $\hat{v}_\rmCoul$ represent
the central, tensor, LS, and Coulomb interactions, respectively.
We define the one-body kinetic energy operator $\hat{t}(x_a)$ and the
two-body potential energy operator $\hat{v}(x_{ab})$ for the later
convenience,
\begin{align}
\hat{t}(x_a) &= -\frac{\hbar^2}{2 M} \frac{A-1}{A} \triangle_a, \\
  \hat{v}(x_{ab}) &= \hat{v}_\rmC(x_{ab})+\hat{v}_\rmT(x_{ab})+\hat{v}_\rmLS(x_{ab})
  +\hat{v}_\rmCoul(x_{ab})+\frac{\hbar^2}{A M}
  \boldsymbol{\nabla}_a\cdot\boldsymbol{\nabla}_b.
\end{align}
The energy correction due to the center of mass motion in
(\ref{eq: hamiltonian1}) is included in $\hat{t}(x_a)$ and
$\hat{v}(x_{ab})$.

We take the expectation value for a Hamiltonian $\hat{H}$ with the
projected wave function and obtain the energy functional,
\begin{align}
 E^{(\pm;Z)} =& \frac{\langle  \Psi^{(\pm;Z)}|\hat{H}
  |\Psi^{(\pm;Z)} \rangle}{\langle  \Psi^{(\pm;Z)} |\Psi^{(\pm;Z)}
 \rangle}
 = \frac{\langle  \Psi^\textrm{intr} |\hat{H}
  | \hat{\mathcal{P}}^\rmps(\pm) \hat{\mathcal{P}}^{\rmc}(Z) \Psi^\textrm{intr} \rangle}
 {\langle \Psi^\textrm{intr}|
  \hat{\mathcal{P}}^\rmps(\pm) \hat{\mathcal{P}}^{\rmc}(Z) \Psi^\textrm{intr}
 \rangle}
  \notag \\
 =& \frac{\frac{1}{4\pi}\int_0^{2 \pi} d \theta
  e^{-iZ\theta} \left(E^{(0)}(\theta)\pm E^{(\rmP)}(\theta)
  \right)}
 {\frac{1}{4\pi}\int_0^{2 \pi} d \theta
  e^{-iZ\theta} \left(n^{(0)}(\theta)\pm n^{(\rmP)}(\theta)
  \right)}.
 \label{eq: Etot}
\end{align}

The denominator in the right-hand side of the above equation is
the normalization of the total wave function,
\begin{align}
 n^{(\pm;Z)} \equiv \langle  \Psi^{(\pm;Z)} |\Psi^{(\pm;Z)} \rangle
 =\frac{1}{4\pi}\int_0^{2 \pi} d \theta
  e^{-iZ\theta} \left(n^{(0)}(\theta)\pm n^{(\rmP)}(\theta)
  \right).
\end{align}
Here, $n^{(0)}(\theta)$ is the determinant of the norm matrix
between the original wave functions $\psi_{\alpha_a}$ and the
charge-rotated wave functions $\psi_{\alpha_a}(\theta)$.
$n^{(\rmP)}(\theta)$ is the determinant of the norm matrix between
the original wave functions $\psi_{\alpha_a}$ and the
parity-inverted and charge-rotated wave functions
$\psi_{\alpha_a}(\theta)$.
\begin{align}
 n^{(0)}(\theta) &\equiv
 \langle \Psi^{\textrm{intr}}| \hat{C}(\theta)| \Psi^{\textrm{intr}} \rangle
 =
 \det B^{(0)}(\theta) \quad
 (B^{(0)}(\theta)_{ab} \equiv \langle
 \psi_{\alpha_a}|\psi_{\alpha_b}(\theta)\rangle)
 ,
 \notag \\
 n^{(\rmP)}(\theta) &\equiv
 \langle \Psi^{\textrm{intr}}|\hat{P}  \hat{C}(\theta) |\Psi^{\textrm{intr}} \rangle
 =
 \det B^{(\rmP)}(\theta) \quad
 (B^{(\rmP)}(\theta)_{ab}
 \equiv \langle \psi_{\alpha_a}|\psi_{\alpha_b}^{(\rmps)}(\theta)\rangle)
 .
\end{align}
The charge-rotated wave function $\psi_{\alpha_a}(x_b;\theta)$ and
the parity-inverted and charge-rotated wave function
$\psi_{\alpha_a}^{(\rmps)}(x_b;\theta)$ are
defined as
\begin{align}
 \psi_{\alpha_a}(x_b;\theta) &\equiv
 e^{i\theta(1+\tau_b^3)/2}  \psi_{\alpha_a}(x_b),
 \\
 \psi_{\alpha_a}^{(\rmps)}(x_b;\theta) &\equiv
 \hat{p}_b e^{i\theta(1+\tau_b^3)/2}  \psi_{\alpha_a}(x_b),
\end{align}
where $\hat{p}_b$ is the single-particle parity operator in
{(\ref{eq: parityop})} and $(1+\tau^3_b)/2$ is the single-particle
proton projection operator in (\ref{eq: chargeop}).

The numerator in the right-hand side of (\ref{eq: Etot}) is
the unnormalized total energy,
\begin{align}
 \langle  \Psi^{(\pm;Z)}|\hat{H}
  |\Psi^{(\pm;Z)} \rangle
 \equiv
 \frac{1}{4\pi}\int_0^{2 \pi} d \theta
  e^{-iZ\theta} \left(E^{(0)}(\theta)\pm E^{(\rmP)}(\theta)
  \right)
  \label{eq: unnormE}
  .
\end{align}
$E^{(0)}(\theta)$ in the right-hand side of (\ref{eq: unnormE})
has a similar form as a simple Hartree-Fock energy but the
single-particle wave functions in the ket are modified by the
charge rotation,
\begin{align}
 &E^{(0)}(\theta)
  \equiv \langle  \Psi^{\textrm{intr}}|\hat{H} \hat{C}(\theta)
 |\Psi^{\textrm{intr}} \rangle
 \notag \\
 &=\sum_{a=1}^A
  \langle \psi_{\alpha_a} |\hat{t}|\tilde{\psi}_{\alpha_a}(\theta)\rangle
  +\sum_{a > b=1}^A
  \langle \psi_{\alpha_a}\psi_{\alpha_b}|\hat{v}|
  \tilde{\psi}_{\alpha_a}(\theta)\tilde{\psi}_{\alpha_b}(\theta) -
  \tilde{\psi}_{\alpha_b}(\theta)\tilde{\psi}_{\alpha_a}(\theta)
  \rangle
  .
\end{align}
Here, $\tilde{\psi}_{\alpha_a}(x;\theta)$ is the superposition of
$\psi_{\alpha_a}(x;\theta)$ weighted by the inverse of the
charge-rotated norm matrix $(B^{(0)}(\theta)^{-1})_{ba}$,
\begin{align}
 \tilde{\psi}_{\alpha_a}(x;\theta)=\sum_{b=1}^A
 \psi_{\alpha_b}(x;\theta)
 (B^{(0)}(\theta)^{-1})_{ba}.
 \label{eq: psitilde}
\end{align}
This summation for $\psi_{\alpha_b}(x;\theta)$ comes from the
antisymmetrization of the total wave function. $E^{(0)}(\theta=0)$
reduces to a simple Hartree-Fock energy. $E^{(\rmP)}(\theta)$ in
the right-hand side of (\ref{eq: unnormE}) has a similar form as
$E^{(0)}(\theta)$ but $\tilde{\psi}_{\alpha_a}(\theta)$'s are
replaced by $\tilde{\psi}_{\alpha_a}^{(\rmps)}(\theta)$'s,
\begin{align}
 &E^{(\rmP)}(\theta)
 \equiv
 \langle
 \Psi^{\textrm{intr}}|\hat{H} \hat{P} \hat{C}(\theta)|
 \Psi^{\textrm{intr}}\rangle
 \notag \\
 &=\sum_{a=1}^A
 \langle \psi_{\alpha_a} |\hat{t}|\tilde{\psi}_{\alpha_a}^{(\rmps)}(\theta)\rangle
 +\sum_{a > b=1}^A
 \langle \psi_{\alpha_a}\psi_{\alpha_b}|\hat{v}|
 \tilde{\psi}_{\alpha_a}^{(\rmps)}(\theta)\tilde{\psi}_{\alpha_b}^{(\rmps)}(\theta) -
 \tilde{\psi}_{\alpha_b}^{(\rmps)}(\theta)\tilde{\psi}_{\alpha_a}^{(\rmps)}(\theta)
 \rangle
 .
\end{align}
Here, $\tilde{\psi}_{\alpha_a}^{(\rmps)}(x;\theta)$ is the sum of
$\psi_{\alpha_a}^{(\rmps)}(x;\theta)$ weighted by the inverse of
the parity-inverted and charge-rotated norm matrix
$(B^{(\rmP)}(\theta)^{-1})_{ba}$,
\begin{align}
 \tilde{\psi}_{\alpha_a}^{(\rmps)}(x;\theta)=\sum_{b=1}^A
 \psi_{\alpha_b}^{(\rmps)}(x;\theta)
 (B^{(\rmP)}(\theta)^{-1})_{ba}.
\end{align}

We then take the variation of $E^{(\pm;Z)}$ with respect to a
single-particle wave function $\psi_{\alpha_a}$,
\begin{align}
 \frac{\delta}{\delta \psi_{\alpha_a}^\dagger (x_a)}
  \left\{E^{(\pm;Z)}-
 \sum_{b,c=1}^A \epsilon_{bc}
 \langle \psi_{\alpha_b} |\psi_{\alpha_c} \rangle
 \right\} = 0
 .
\end{align}
The Lagrange multiplier $\epsilon_{ab}$ is introduce to guarantee
the ortho-normalization of a single-particle wave function,
$\langle \psi_{\alpha_a}|\psi_{\alpha_b}\rangle
=\delta_{\alpha_a,\alpha_b}$. As the result, we obtain the
following Hartree-Fock-like equation with the charge and parity
projections (the CPPHF equation) for each $\psi_{\alpha_a}$,
\begin{align}
 \frac{1}{4\pi}\int_0^{2 \pi} d \theta e^{-iZ\theta}
 \Biggl[
 &
 n^{(0)}(\theta)
 \biggl\{\hat{t}(x_a) \tilde{\psi}_{\alpha_a}(x_a;\theta)
 +
 \sum_{b=1}^A \langle \psi_{\alpha_b}|
 \hat{v}(x_{a1})|\tilde{\psi}_{\alpha_b}(\theta) \rangle_1
 \tilde{\psi}_{\alpha_a}(x_a;\theta)
 \notag \\
 &
 -\sum_{b=1}^A \langle \psi_{\alpha_b}|
 \hat{v}(x_{a1})|\tilde{\psi}_{\alpha_a}(\theta) \rangle_1
 \tilde{\psi}_{\alpha_b}(x_a;\theta)
 \notag \\
 &
 -(E^{(\pm;Z)}-E^{(0)}(\theta))
 \tilde{\psi}_{\alpha_a}(x_a;\theta)
 -\sum_{b=1}^A\eta^{(0)}_{ba}(\theta) \tilde{\psi}_{\alpha_b}(x_a;\theta)
 \biggr\}
 \notag\\
 &\pm
 n^{(\rmP)}(\theta)
 \biggl\{\hat{t}(x_a) \tilde{\psi}_{\alpha_a}^{(\rmps)}(x_a;\theta)
 +
 \sum_{b=1}^A \langle \psi_{\alpha_b}|\hat{v}(x_{a1})|\tilde{\psi}_{\alpha_b}^{(\rmps)}(\theta) \rangle_1
 \tilde{\psi}_{\alpha_a}^{(\rmps)}(x_a;\theta)
 \notag \\
 &
 -\sum_{b=1}^A \langle \psi_{\alpha_b}|\hat{v}(x_{a1})|\tilde{\psi}_{\alpha_a}^{(\rmps)}(\theta) \rangle_1
 \tilde{\psi}_{\alpha_b}^{(\rmps)}(x_a;\theta)
 \notag \\
 &
 -(E^{(\pm;Z)}-E^{(\rmP)}(\theta))
 \tilde{\psi}_{\alpha_a}^{(\rmps)}(x_a;\theta)
 -\sum_{b=1}^A
 \eta^{(\rmP)}_{ba}(\theta)
 \tilde{\psi}_{\alpha_b}^{(\rmps)}(x_a;\theta)
 \biggr\}
 \Biggr]
 \notag\\
 &=n^{(\pm;Z)} \sum_{b=1}^A \epsilon_{ab} \psi_{\alpha_b}(x_a)
 ,
 \label{eq: cphf}
\end{align}
where $a=1,2,\dots,A$.
Here, $\eta_{ab}^{(0)}(\theta)$ and
$\eta_{ab}^{(\rmP)}(\theta)$ are defined as follows,
  \begin{align}
  \eta_{ab}^{(0)}(\theta)
  \equiv
  &
  \langle \psi_{\alpha_a} |\hat{t}|\tilde{\psi}_{\alpha_b}(\theta)\rangle
  \notag \\
  &+\sum_{c=1}^A
  \langle \psi_{\alpha_a}\psi_{\alpha_c}|\hat{v}|
  \tilde{\psi}_{\alpha_b}(\theta)\tilde{\psi}_{\alpha_c}(\theta) -
  \tilde{\psi}_{\alpha_c}(\theta)\tilde{\psi}_{\alpha_b}(\theta)
  \rangle
  ,
  \\
  \eta_{ab}^{(\rmP)}(\theta)
  \equiv
  &
  \langle \psi_{\alpha_a} |\hat{t}|\tilde{\psi}_{\alpha_b}^{(\rmps)}(\theta)\rangle
  \notag \\
  &+\sum_{c=1}^A
  \langle \psi_{\alpha_a}\psi_{\alpha_c}|\hat{v}|
  \tilde{\psi}_{\alpha_b}^{(\rmps)}(\theta)\tilde{\psi}_{\alpha_c}^{(\rmps)}(\theta) -
  \tilde{\psi}_{\alpha_c}^{(\rmps)}(\theta)\tilde{\psi}_{\alpha_b}^{(\rmps)}(\theta)
  \rangle,
  \end{align}
and the notation for the integration of the two-body matrix elements,
\begin{align}
 \langle \psi_{\alpha_b}|\hat{v}(x_{a1})|\psi_{\alpha_c} \rangle_{1}
  =\int dx_1
  \psi^\dagger_{\alpha_b}(x_1) \hat{v}(x_{a1})
  \psi_{\alpha_c}(x_1).
\end{align}
The system of the coupled equations (\ref{eq: cphf}) for
$a=1,\cdots,A$ is solved self-consistently. We note here that the
CPPHF equation reduces to the parity-projected Hartree-Fock
equation with only the parity projection by setting $\theta=0$ in
(\ref{eq: cphf}), which was already obtained by S.~Takami
\textit{et al.} \cite{takami96}.

We give here the expressions for the expectation value of the
kinetic energy $\langle \hat{T} \rangle^{(\pm;Z)}$ with the center
of mass correction and that of the two-body potential energy
$\langle \hat{v}_\sigma \rangle^{(\pm;Z)}$ for
$\hat{v}_\sigma$ ($\sigma=\rmC$, $\rmT$, $\rmLS$, and $\rmCoul$)
for the later convenience.
\begin{align}
 \langle \hat{T} \rangle^{(\pm;Z)}
 =&
 \frac{1}{4 \pi n^{(\pm;Z)}} \int_0^{2 \pi} d \theta e^{-i Z \theta}
 \Biggl[
 n^{0}(\theta)\Biggl\{
 \sum_{a=1}^A
 \langle \psi_{\alpha_a} |\hat{t}|\tilde{\psi}_{\alpha_a}(\theta)\rangle
 \notag \\
 &+\sum_{a > b=1}^A
 \langle \psi_{\alpha_a}\psi_{\alpha_b}|\frac{\hbar^2}{A M}
 \boldsymbol{\nabla}_a\cdot\boldsymbol{\nabla}_b|
 \tilde{\psi}_{\alpha_a}(\theta)\tilde{\psi}_{\alpha_b}(\theta) -
 \tilde{\psi}_{\alpha_b}(\theta)\tilde{\psi}_{\alpha_a}(\theta)
 \rangle
 \Biggr\}
 \notag\\
 &\pm
 n^{(\rmP)}(\theta)
 \Biggl\{\sum_{a=1}^A
 \langle \psi_{\alpha_a}
 |\hat{t}|\tilde{\psi}_{\alpha_a}^{(\rmps)}(\theta)\rangle
 \notag \\
 &
 +\sum_{a > b=1}^A
 \langle \psi_{\alpha_a}\psi_{\alpha_b}|\frac{\hbar^2}{A M}
 \boldsymbol{\nabla}_a\cdot\boldsymbol{\nabla}_b|
 \tilde{\psi}_{\alpha_a}^{(\rmps)}(\theta)\tilde{\psi}_{\alpha_b}^{(\rmps)}(\theta) -
 \tilde{\psi}_{\alpha_b}^{(\rmps)}(\theta)\tilde{\psi}_{\alpha_a}^{(\rmps)}(\theta)
 \rangle
 \Biggr\}
 \Biggr],
 \\
 \langle \hat{v}_\sigma \rangle^{(\pm;Z)}
 =&
 \frac{1}{4 \pi n^{(\pm;Z)}} \int_0^{2 \pi} d \theta e^{-i Z \theta}
 \notag \\
 &\times \Biggl\{
 n^{0}(\theta)
 \sum_{a > b=1}^A
 \langle \psi_{\alpha_a}\psi_{\alpha_b}|\hat{v}_\sigma|
 \tilde{\psi}_{\alpha_a}(\theta)\tilde{\psi}_{\alpha_b}(\theta) -
 \tilde{\psi}_{\alpha_b}(\theta)\tilde{\psi}_{\alpha_a}(\theta)
 \rangle
 \notag\\
 &
 \pm
 n^{(\rmP)}(\theta)
 \sum_{a > b=1}^A
 \langle \psi_{\alpha_a}\psi_{\alpha_b}|\hat{v}_\sigma|
 \tilde{\psi}_{\alpha_a}^{(\rmps)}(\theta)\tilde{\psi}_{\alpha_b}^{(\rmps)}(\theta) -
 \tilde{\psi}_{\alpha_b}^{(\rmps)}(\theta)\tilde{\psi}_{\alpha_a}^{(\rmps)}(\theta)
 \rangle
 \Biggr\}.
\end{align}
\section{Application to the alpha particle} \label{sec: result}
We apply the CPPHF method to the alpha particle (A=4, Z=2), which
has been studied extensively with various theoretical methods
including exact calculations. As the ground state configuration of
the alpha particle, we assume four states $n=1,2$ and $(jm)=(1/2,
\pm 1/2)$, which are fully occupied. Here, $j$ and $m$ are the
total angular momentum and the magnetic quantum number of a
single-particle state. The index $n$ labels the two different
states which have the same $j$ and $m$. The intrinsic wave
function (\ref{eq: wf intr}) for the alpha particle becomes
\begin{align}
 \Psi^{\textrm{intr}} = \hat{\mathcal{A}} \prod_{n=1,2}
 \prod_{m=\pm \frac{1}{2}} \psi_{nj=\frac{1}{2}m}(x).
 \label{eq: wf intr alpha}
\end{align}
Because we fill $m$=$\pm1/2$ states with $j=1/2$ for each $n$,
$\Psi^{\textrm{intr}}$ has the total angular momentum 0. For each
$n$, $\psi_{nj=\frac{1}{2}m}$ becomes
\begin{align}
 {\psi_{nj=\frac{1}{2}m}}(x) =
  \sum_{t_z=\pm \frac{1}{2}}\bigl(
             &{
  \phi_{n l=0 t_z}(r) \mathcal{Y}_{j=\frac{1}{2} l=0 m}(\Omega)\zeta(t_z)}
  \notag \\
  +&{
  \phi_{n \bar{l}=1 t_z}(r) \mathcal{Y}_{j=\frac{1}{2} \bar{l}=1 m}(\Omega)
  \zeta(t_z)}
  \bigr)
  .
 \label{eq: spwf alpha}
\end{align}
The single-particle wave function $\psi_{nj=\frac{1}{2}m}$
consists of four components, which have different parities $(\pm)$
and different charges $(t_z=\pm 1/2)$. The first term on the
right-hand-side has $j=1/2$ and $l=0$ (a $s$-state with
positive-parity) and the second term has $j=1/2$ and $l=1$ (a
$p$-state with negative-parity). The single-particle $p$-state
probability $P(-)$, which is a measure for how much the $p$-state
mixes into the simple $(0s)^4$ configuration in the intrinsic
state of the alpha particle, is defined as
\begin{align}
 P(-) = \frac{1}{2} \sum_{n=1,2} \sum_{t_z=\pm \frac{1}{2}}
 \int d r r^2 \phi_{n l=1 t_z}^\dagger(r) \phi_{n l=1t_z} (r).
 \label{eq: p-prob}
\end{align}
The factor 1/2 in front of the right hand side of (\ref{eq:
p-prob}) is added to average the single-particle $p$-state
probability over the states for $n=1,2$. The four-body intrinsic
wave function (\ref{eq: wf intr alpha}) does not have a good
parity and a definite charge number; it is a  mixture of positive
and negative parities, and of various charge numbers Z=0$\sim$4.
In the alpha particle, the ground state has a positive parity and
the charge number Z=2. Therefore, we act the projection operator
$\hat{\mathcal{P}}^\rmps (+)\hat{\mathcal{P}}^\rmc(2)$ on the
intrinsic wave function (\ref{eq: wf intr alpha}).

In the numerical calculation, we expand the radial wave function
$\phi$ in (\ref{eq: spwf alpha}) by the Gaussian basis with the
geometric-series widths \cite{hiyama03}. We take the number of
basis as 10 and set the minimum width to 0.5 fm and the maximum
width to 6.0 fm. To solve the CPPHF equation~(\ref{eq: cphf})
self-consistently we use the gradient method
\cite{takami96,reinhard91}.

For the central force, we take the Volkov force No.~1
\cite{volkov65} as a reference and introduce the multiplying
factor $x_\rmTE$ for the triplet-even part.
\begin{align}
 \hat{v}^{x_{\rmTE}}_\rmC (x_{ab})
 =&-v_\mathrm{A} \exp(-(r_{ab}/\alpha_\mathrm{A})^2)
 \left(
 x_{\rmTE} P^{^3\rmE}_{ab}+P^{^1\rmE}_{ab}+(1-2m_\rmV)(P^{^3\rmO}_{ab}+P^{^1\rmO}_{ab})
 \right)
 \notag \\
 &
 +v_\mathrm{R} \exp(-(r_{ab}/\alpha_\mathrm{R})^2)
 \left(
 P^{^3\rmE}_{ab}+P^{^1\rmE}_{ab}+(1-2m_\rmV)(P^{^3\rmO}_{ab}+P^{^1\rmO}_{ab})
 \right)
 .
  \label{eq: volkov xte}
\end{align}
Here, $P^{^3\rmE}_{ab}$, $P^{^1\rmE}_{ab}$, $P^{^3\rmO}_{ab}$, and
$P^{^1\rmO}_{ab}$ are the projection operators for the
triplet-even, singlet-even, triplet-odd, and singlet-odd states,
respectively. If we set $x_\rmTE = 1$, $\hat{v}^{x_{\rmTE}}_\rmC$
reduces to the original Volkov force. The term proportional to
$v_\mathrm{A}$ is a middle-range attractive part and that to
$v_\mathrm{R}$ is a short-range repulsive one. The values of the
parameters are $v_\mathrm{A}=83.34$ MeV, $\alpha_\mathrm{A}=1.6$
fm, $v_\mathrm{R}=144.86$ MeV, and $\alpha_\mathrm{R}=0.82$ fm. We
fix the Majorana parameter $m_\rmV =0.6$. The Volkov force
reproduces the binding energy of the alpha particle without
non-central forces. We can consider that the effect of the tensor
force is included in the Volkov force by renormalizing the central
force. This renormalization affects the middle range of the $^3$E
part of the central force mainly. Therefore, we reduce the $^3$E
part of the central force in the Volkov force by multiplying the
factor $x_{\rmTE}$ to the attractive $^3$E part of the Volkov
interaction, when we include the tensor force explicitly.

As for the non-central forces (the tensor and the LS forces), the
G3RS \cite{tamagaki68} force is adopted as a reference, which is
determined so as to reproduce the scattering phase shift of NN
scattering and have the one-pion-exchange potential tail in the
tensor part. In the mean-field-type framework, we have to treat
the effect of the short-range correlation on the tensor
interaction and further take into account the effect of the
$\Delta$ isobar-hole excitations \cite{oset82}, separately from
the mean-field correlation. In the present study, we shall treat
these effects by multiplying a factor $x_{\rmT}$ to the
$\boldsymbol{\tau}_1\cdot\boldsymbol{\tau}_2$ part of the tensor
force, which is the dominant part of the tensor force and mediated
mainly by the pion. We need further studies on this factor in the
near future. The tensor part of the G3RS force with the
multiplying factor $x_{\rmT}$ can be written as following,
\begin{align}
 &\hat{v}^{x_\rmT}_\rmT(x_{ab})
 =\frac{1}{4}
 \Biggl[
 \Biggl\{
 \sum_{n=1}^3 v_{\rmT n}^{^3\rmE} \exp(-(r_{ab}/\eta_{\rmT n}^{^3\rmE})^2)
  +3 \sum_{n=1}^3 v_{\rmT n}^{^3\rmO} \exp(-(r_{ab}/\eta_{\rmT n}^{^3\rmO})^2)
  \Biggr\}
 \notag \\
 &
 + x_\rmT \Biggl\{
 -\sum_{n=1}^3 v_{\rmT n}^{^3\rmE} \exp(-(r_{ab}/\eta_{\rmT n}^{^3\rmE})^2)
 +3 \sum_{n=1}^3 v_{\rmT n}^{^3\rmO} \exp(-(r_{ab}/\eta_{\rmT n}^{^3\rmO})^2)
 \Biggr\} \boldsymbol{\tau}_a\cdot\boldsymbol{\tau}_b
 \Biggr]S^T_{ab}
  ,
  \label{eq: g3rs_xt}
\end{align}
where the tensor operator $S^\rmT_{ab}$ is defined as,
\begin{align}
 S^\rmT_{ab} = 3 \frac{(\boldsymbol{\sigma}_a\cdot\boldsymbol{r}_{ab})
  (\boldsymbol{\sigma}_b\cdot\boldsymbol{r}_{ab})}{r_{ab}^2}
  -\boldsymbol{\sigma}_a\cdot\boldsymbol{\sigma}_b
  .
\end{align}
The factors, $x_{\rmTE}$ and $x_{\rmT}$ are correlated and are
determined so as to reproduce the binding energy of the alpha
particle.

The LS part of the G3RS force is made from the two-range
Gaussians:
\begin{align}
 \hat{v}_{\rmLS}(x_{ab}) =&
 \sum_{n=1}^2 v_{\rmLS n}^{^3\rmO} \exp(-(r_{ab}/\eta_{\rmLS n}^{^3\rmO})^2)P^{^3\rmO}
  \boldsymbol{L}_{ab}\cdot\boldsymbol{S}_{ab}
  .
  \label{eq: g3rs_LS}
\end{align}

The parameters of the G3RS force used in this work is summarized in
Table~\ref{table: g3rs}.
\begin{table}[bt]
\begin{center}
\caption{The parameters for the G3RS force used in this
 work \cite{tamagaki68}. The second, the third, and the fourth
 columns are the interaction strengths $v$ in MeV and the corresponding ranges $\eta$
 in fm for the $^3$E tensor and $^3$O tensor and $^3$O LS channels, respectively.
 (See Eqs.~(\ref{eq: g3rs_xt}) and (\ref{eq: g3rs_LS}).)
 \label{table: g3rs}}
\begin{tabular}{crrr}
\hline
 & \multicolumn{2}{c}{Tensor} & \multicolumn{1}{c}{LS}\\
 & $^3\rmE$ & $^3\rmO$ & $^3\rmO$ \\
 \hline
 $v_1$ (MeV) & -7.5 & 2.5 & -800 \\
 $\eta_1$ (fm) &2.5 & 2.5 & 0.6 \\
 $v_2$ (MeV)& -67.5 & 20. & 800 \\
 $\eta_2$ (fm) & 1.2 & 1.2 & 0.4 \\
 $v_3$ (MeV)&  67.5 & -20. & \\
 $\eta_3$ (fm) & 0.447 & 0.6 & \\
 \hline
\end{tabular}
\end{center}
\end{table}
We take further the Coulomb interaction,
\begin{align}
 \hat{v}_{\rmCoul}(x_{ab})= \frac{e^2}{r_{ab}},
\end{align}
with $e$ being the charge of the electron.

\begin{table}[bt]
\begin{center}
\caption{The results for the ground ($0^+$) state of the alpha
particle for
 various cases. HF denotes the simple Hartree-Fock scheme. PPHF denotes the
 parity-projected Hartree-Fock scheme in which only the
 parity-projection is performed. CPPHF denotes the charge- and
 parity-projected Hartree-Fock scheme in which both the charge and
 parity projections are performed.
 The potential energy ($\langle \hat{v} \rangle^{(+;2)}$),
 kinetic energy ($\langle \hat{T} \rangle^{(+;2)}$), total energy
 ($E^{(+;2)}$), root-mean-square matter radius ($R_\rmm$),
 and the probability of the $p$-state component (P(-)) are given in the table.
 $\langle \hat{v}_\rmC \rangle^{(+;2)}$, $\langle \hat{v}_{\rmT} \rangle^{(+;2)}$,
 $\langle \hat{v}_{\rmLS} \rangle^{(+;2)}$, and $\langle \hat{v}_{\rmCoul} \rangle^{(+;2)}$
 are the expectation values for the central, tensor, LS, and Coulomb potentials,
 respectively.  The factors, $x_\rmT$ and $x_{\rmTE}$, denote the factors multiplied
 to the $\boldsymbol{\tau}_1\cdot\boldsymbol{\tau}_2$-type tensor force and the $^3$E central force.
 \label{table: cphf}}
\begin{tabular}{crrrr}
\hline
& \multicolumn{1}{c}{HF}&  \multicolumn{1}{c}{HF}&
 \multicolumn{1}{c}{PPHF}&\multicolumn{1}{c}{CPPHF}\\
$x_\rmT$&      1.0&    1.5&        1.5&        1.5\\
$x_{\rmTE}$&   1.0&    0.81&   0.81&       0.81\\
\hline
$\langle \hat{v}_\rmC \rangle^{(+;2)}$ (MeV)&   -76.67&       -56.85&   -61.31&     -64.75\\
$\langle \hat{v}_\rmT \rangle^{(+;2)}$ (MeV)&   0.00&  0.00&    -10.91&     -30.59\\
$\langle \hat{v}_{\rmLS} \rangle^{(+;2)}$ (MeV)&   0.00&   0.00&      0.67&       1.91\\
$\langle \hat{v}_{\rmCoul} \rangle^{(+;2)}$ (MeV)&   0.83&     0.76&      0.78&       0.85\\
$\langle \hat{v} \rangle^{(+;2)}$ (MeV)& -75.84&       -56.10&   -70.76&     -92.58\\
$\langle \hat{T} \rangle^{(+;2)}$ (MeV)&  48.54&        39.98&      49.67&      64.39\\
$E^{(+;2)}$ (MeV)&-27.30&       -16.12&         -21.09&         -28.19\\
$R_\rmm$ (fm)& 1.48&        1.63&        1.51&        1.39\\
$P(-)$ (\%)&  0.0&        0.0&        7.6&        16.1\\
\hline
\end{tabular}
\end{center}
\end{table}

We present here the results of the CPPHF calculations for the
alpha particle. To see the effect of the charge and parity
mixings, we calculate three cases. The first calculation is the
simple Hartree-Fock (HF) scheme, in which we do not perform neither
the parity nor charge projection. The second calculation is the
parity-projected Hartree-Fock (PPHF) scheme, in which we only
perform the parity projection. The third calculation is the
charge- and parity-projected Hartree-Fock (CPPHF) scheme, in which
we perform both the parity and charge projections. We show the
calculated results in Table~\ref{table: cphf}, where we take
$x_{\rmT}=1.5$ and $x_{\rmTE}=0.81$, with which the binding energy
of the alpha particle is reproduced in the CPPHF case. For
comparison, we show the results for the HF calculation with the
original Volkov No.~1 force ($x_\rmT=1.0$ and $x_{\rmTE}=1.0$).

In the simple HF case the energy from the tensor force $\langle
\hat{v}_\rmT \rangle^{(+;2)}$ is zero. It means that the result of
the HF calculation simply becomes a $(0s)^4$ configuration and
there is no $p$-state component. In this case the expectation
value of the tensor force is zero identically, because the tensor
force does not act between $s$-states. If we perform the
parity-projection (PPHF), the energy contribution from the tensor
force becomes finite. The kinetic energy becomes larger because
some component of the $s$-state is shifted up to $p$-states to
gain the correlation caused by the tensor force. We perform then
the charge projection (CPPHF) further.  We see the contribution
from the tensor force becomes much larger. It is reasonable
because in the PPHF case only the $\tau^0_1\tau^0_2$ component of
$\boldsymbol{\tau}_1\cdot\boldsymbol{\tau}_2$ in the tensor force
is active, while in the CPPHF case all the $\tau^+_1\tau^-_2$,
$\tau^0_1\tau^0_2$, and $\tau^-_1\tau^+_2$ parts of the tensor
force are active.  In fact, the energy from the tensor force in
the CPPHF case is about three times large than that in the PPHF
case. The results indicate that both the parity and charge
projections are very important to treat the tensor force in a
mean-field-type model.

We note that the variation-after-projection scheme that we take
here, is needed because even if we assume the mixing of parity and
charge in the simple Hartree-Fock calculation, we cannot obtain
the result with mixed symmetries. The parity mixed state contains
odd parity states, which require large energy jump across a major
shell. In fact, the energy gap between $0s$-shell and $0p$-shell
in the alpha particle is more than 20 MeV. Therefore, we need to
perform the parity projection before variation to remove the
spurious component mixing into the intrinsic wave function
$\Psi^{\textrm{intr}}$ in (\ref{eq: wf intr alpha}) properly. In
contrast to the parity mixing, the correlations by deformation and
pairing can be treated quite nicely with the
projection-after-variation scheme including a configuration mixing
along the quadrupole degree of freedom
\cite{valor00,rodriguez-guzman00,rodriguez-guzman02,bender03},
where the rotation symmetry and the particle-number symmetry are
broken in a intrinsic state. The correlations by deformation and
pairing are in principle those within the same major shell and
causes small energy loss and therefore the mixing of the angular
momentum and the particle number occur even in the
projection-after-variation scheme.

\begin{table}[bt]
\begin{center}
\caption{
 The results for the ground ($0^+$) state of the alpha particle with the CPPHF scheme.
 We change $x_\rmT$ from 1.0 to 2.0 and determine $x_{\rmTE}$ accordingly
 so as to reproduce the binding energy of the alpha particle.
 The potential energy ($\langle \hat{v} \rangle^{(+;2)}$),
 kinetic energy ($\langle \hat{T} \rangle^{(+;2)}$), total energy
 ($E^{(+;2)}$), root-mean-square matter radius ($R_\rmm$),
 root-mean-square charge radius ($R_\rmc$),
 the probability of the $p$-state component ($P(-)$),
 and the probability of the $S=2$ component ($P(S=2)$) are shown.
 $\langle \hat{v}_\rmC \rangle^{(+;2)}$, $\langle \hat{v}_{\rmT} \rangle^{(+;2)}$,
 $\langle \hat{v}_{\rmLS} \rangle^{(+;2)}$, and $\langle \hat{v}_{\rmCoul} \rangle^{(+;2)}$
 are the expectation values for the central, tensor, LS, and Coulomb potentials, respectively.
\label{table: tte}}
\begin{tabular}{crrrrr}
 \hline
 $x_\rmT$&            1&       1.25&        1.5&       1.75&        2\\
 $x_{\rmTE}$&     0.93&      0.88&     0.81&   0.73&     0.64\\
 \hline
 $\langle \hat{v}_\rmC \rangle^{(+;2)}$ (MeV)&  -73.60&    -70.13&     -64.75&     -58.34&     -50.69\\
$\langle \hat{v}_\rmT \rangle^{(+;2)}$ (MeV)&-12.26&   -20.23&     -30.59&     -43.86&     -60.41\\
$\langle \hat{v}_{\rmLS} \rangle^{(+;2)}$ (MeV)&0.75&  1.26&   1.91&   2.72&   3.72\\
$\langle \hat{v}_{\rmCoul} \rangle^{(+;2)}$ (MeV)&0.85&    0.85&   0.85&   0.86&   0.87\\
$\langle \hat{v} \rangle^{(+;2)}$ (MeV)&-84.27&     -88.26&     -92.58&     -98.62&     -106.51\\
$\langle \hat{T} \rangle^{(+;2)}$ (MeV)&55.81&  59.71&  64.39&  70.52&  78.19\\
$E^{(+;2)}$ (MeV)&-28.46&   -28.55&     -28.19&     -28.10&     -28.32\\
$R_\rmm$ (fm) &1.43&    1.41&    1.39&    1.36&    1.33\\
$R_\rmc$ (fm) &1.64&    1.62&    1.60&    1.58&    1.56\\
$P(-)$ (\%)&   10.5&   13.3&   16.1&   18.8&   21.4\\
$P(S=2)$ (\%)& 2.9     &4.9     &7.3     &10.1    &13.3\\
%$\langle\Psi^\textrm{intr}|P|\Psi^\textrm{intr}
%\rangle/\langle\Psi^\textrm{intr}|\Psi^\textrm{intr}\rangle$&
% 0.74&     0.68&   0.63&   0.58&   0.54\\
\hline
\end{tabular}
\end{center}
\end{table}
We check now the dependence of the results on the strength of the
tensor force by changing $x_{\rmT}$ from 1.0 to 2.0. We determine
$x_{\rmTE}$ so as to reproduce the binding energy of the alpha
particle for each $x_\rmT$. We show the results of the CPPHF
scheme in Table~\ref{table: tte}. From the table we can see that
the contribution of the energy from the tensor force becomes
larger with $x_{\rmT}$. The kinetic energy becomes larger also.
The probability of the $p$-state (P(-)) changes from 10.5\% to
21.4\%. In the last row in Table~\ref{table: tte}, we show the
probability of the total spin $S=2$ component $P(S=2)$, which is
defined as,
\begin{align}
P(S=2) = \sum_{M_s=-2}^2 \frac{\langle \Psi^{(+;2)}|
\Phi_{S=2}(M_S) \rangle \langle \Phi_{S=2}(M_S)  |\Psi^{(+;2)}
\rangle}{\langle \Psi^{(+;2)}|\Psi^{(+;2)}\rangle},
\end{align}
where
\begin{align}
\Phi_{S=2}(M_S)=[[\chi_1 \times \chi_2]^{(1)} \times [\chi_3
\times \chi_4]^{(1)}]^{(2)}_{M_S}.
\end{align}
$P(S=2)$ corresponds to the $D$-state probability, because in the
present calculation the total wave function has the $0^+$
spin-parity. $P(S=2)$ changes from 2.9\% to 13.3\% when $x_\rmT$
changes from 1.0 to 2.0, and is almost proportional to the tensor
correlation energy $\langle \hat{v}_\rmT \rangle^{(+;2)}$. We note
that the $D$-state probability of the alpha particle in the recent
exact-type calculations \cite{kamada01} is about 15\%. The matter
root-mean-square radius $R_\rmm$ decreases with $x_\rmT$. The
root-mean-square charge radius $R_\rmc$ is calculated from the
proton root-mean-square radius $R_\rmps$ as $R_\rmc =
\sqrt{R_\rmps^2+0.64}$. This approximation for $R_\rmc$
corresponds to assuming the charge radius of proton as 0.80 fm.
The charge radii obtained here are slightly smaller than the
experimental value $R_\rmc = 1.676(8)$ fm \cite{vries87}. The
charge form factor has a dip around the momentum transfer squared
$q^2=15$ fm$^{-2}$ for the case without the tensor force, and with
the inclusion of the tensor force the position of the dip moves
towards smaller momentum transfer. The position of the dip is
around $q^2=11$ fm$^{-2}$ for $x_\rmT=1.5$ in accordance with the
experiment, but the amount of the second bump is somewhat
underestimated.

\begin{figure}[h]
 \centerline{
 \includegraphics[width=10cm]{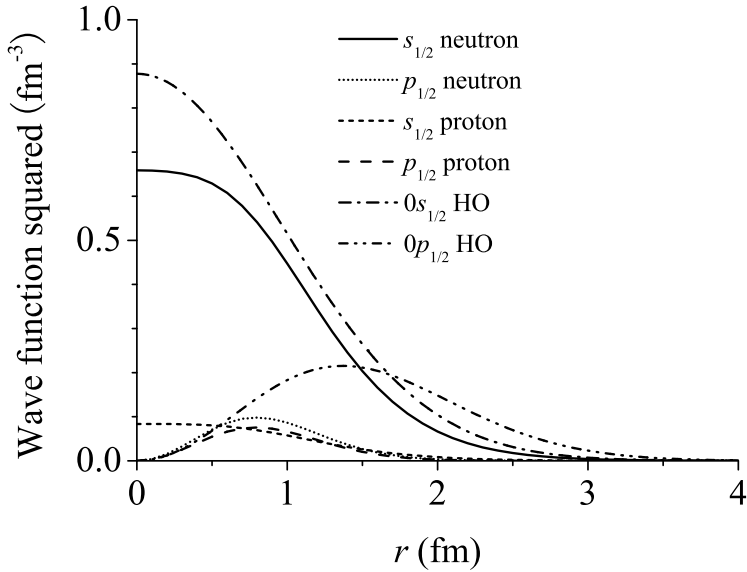}}
 \caption{Single particle wave functions squared for the case with
 $x_{\rmT}=1.5$ and
 $x_{\rmTE}=0.81$ as a function of the radial coordinate $r$ (fm).
 The solid curve denotes the positive-parity neutron
 component $|\phi_{10-\frac{1}{2}}|^2$ ($l=0$),
 the dotted curve the negative-parity neutron
 component $|\phi_{11-\frac{1}{2}}|^2$ ($l=1$),
 the short-dashed curve the positive-parity proton
 component $|\phi_{10\frac{1}{2}}|^2$ ($l=0$), and
 the long-dashed curve the negative-parity proton
 component $|\phi_{11\frac{1}{2}}|^2$ ($l=1$).
 For comparison, we show the harmonic oscillator (HO) wave functions with the oscillator length 1.37
 fm for the $0s$ state (dashed and dotted) and the $0p$ state
 (dashed and double dotted).
 }
 \label{fig: wf}
\end{figure}
In Fig.~\ref{fig: wf}, we show one of the intrinsic
single-particle wave function (squared wave function) for the
alpha particle with $x_{\rmT}=1.$5 in the CPPHF method. The
dominant component of this wave function is the $s$-state
(positive-parity state) and neutron. Due to the charge and parity
mixings, there appear the components of the $p$-state and neutron,
$s$-state and proton, and $p$-state and proton. In the wave
function the probability of the negative parity component is 16\%
and that of the proton component is 17\%. There is one more
intrinsic single-particle state ($n=2$), which has the
$s$-state-and-proton component as the dominant one. In the figure
the harmonic oscillator (HO) wave functions with the oscillator
length $b = 1.37$ fm for the $0s$ and $0p$ state are also plotted
for comparison. The $0s$-state HO wave function almost corresponds
to the calculated wave function in the simple HF scheme with the
original Volkov interaction ($x_\rmT=1.0$, $x_\rmTE=1.0$). The
$s$-state and neutron component in the CPPHF method has almost the
same width as the HO wave function but the magnitude is somehow
reduced. This reduction is caused by the parity and charge
mixings. The mixing of parity reduces the wave function around the
origin. It is interesting to see that the negative parity
components have narrower widths as compared to the $0p$-state HO
wave function \cite{akaishi02}. If we calculate the overlap
between the $p$-state wave function in the CPPHF method with the
$0p$-state HO wave function changing the oscillator length
$b_{0p}$, the maximum overlap is achieved at $b_{0p}=0.81$ fm. It
means that the negative parity component in the CPPHF method is
not a simple $0p$ state. The mixing of a higher momentum component
is necessary to make the $p$-state have such a compact
distribution. This fact suggests that the tensor force induces
higher momentum components, which are not included in a simple
shell model configuration. The mixing of the higher-momentum
component results in the increase of the kinetic energy as seen in
Table~\ref{table: tte}. The correlation induced by the tensor
force produces more attractive energy and therefore the mixing of
the higher-momentum component is favorable in total.

\begin{table}[h]
\begin{center}
\caption{The results for the first $0^-$ state in the alpha
particle. The potential energy ($\langle \hat{v}
\rangle^{(-;2)}$),
 kinetic energy ($\langle \hat{T} \rangle^{(-;2)}$), total energy
 ($E^{(-;2)}$), root-mean-square matter radius ($R_\rmm$), and
 the probability of the $p$-state component ($P(-)$) are shown.
 $\langle \hat{v}_\rmC \rangle^{(-;2)}$, $\langle \hat{v}_{\rmT} \rangle^{(-;2)}$,
 $\langle \hat{v}_{\rmLS} \rangle^{(-;2)}$, and $\langle \hat{v}_{\rmCoul} \rangle^{(-;2)}$
 are the expectation values for the central, tensor, LS, and Coulomb potentials, respectively.
\label{table: alpha_neg}}
\begin{tabular}[t]{crrrrr}
\hline
$x_\rmT$   &1  &1.25   &1.5    &1.75   &2 \\
$x_{\rmTE}$    &0.93   &0.88   &0.81   &0.73   &0.64 \\
 \hline
 $\langle \hat{v}_\rmC \rangle^{(-;2)}$ (MeV)  &-29.05 &-29.73 &-29.36 &-28.16 &-26.03 \\
 $\langle \hat{v}_\rmT \rangle^{(-;2)}$    &-4.57  &-8.68  &-14.00 &-20.49 &-28.18 \\
 $\langle \hat{v}_{\rmLS} \rangle^{(-;2)}$     &0.38   &0.73   &1.11   &1.52   &1.96 \\
 $\langle \hat{v}_{\rmCoul} \rangle^{(-;2)}$ &0.39     &0.48   &0.54   &0.58   &0.61 \\
 $\langle \hat{v} \rangle^{(-;2)}$  &-32.85 &-37.21 &-41.72 &-46.55 &-51.63 \\
 $\langle \hat{T} \rangle^{(-;2)}$ &30.15   &34.43  &38.99  &43.80  &48.80 \\
$E^{(-;2)}$ &-2.70  &-2.78  &-2.73  &-2.76  &-2.83 \\
$R_\rmm$ (fm)  &3.23&    2.74&    2.40&    2.19&    2.06\\
$P(-)$ (\%) &10.7 &   15.1 &   19.3 &   22.8 &   25.8\\
\hline
\end{tabular}
\end{center}
\end{table}
It is an interesting subject to solve the $0^-$ state in the CPPHF
method, since the $0^-$ state is the daughter state of the $0^+$
ground state in the parity- and charge-mixed intrinsic state. We
calculate the $0^-$ state of the alpha particle with the same
parameters as the $0^+$ state. The results are tabulated in
Table~\ref{table: alpha_neg}.  The attraction due to the tensor
force increases with the tensor parameter $x_{\rmT}$, but the
amount is less than a half of the case of the positive parity
state.  The net binding energy comes out to be about 2.8 MeV.  It
is interesting to note, however, that the admixture of the
opposite parity components $P(-)$ in the single-particle states is
larger than the case of the $0^+$ state as seen in
Table~\ref{table: tte}. This is related with the fact that $P(-)$
ought to be finite for the $0^-$ state even for the case without
the tensor force, while for the $0^+$ case $P(-)$ becomes zero in
such a case. From Table~\ref{table: tte} and \ref{table:
alpha_neg} the excitation energy $E_x$ from the ground state of
the $0^-$ state in the CPPHF method is about 25.6 MeV in our
calculation. The experimental value for the first $0^-$ state,
which corresponds to the $0^-$ state calculated here, is $E_x
=21.010$ MeV with the total width $\Gamma =  0.840$ MeV
\cite{he4level}. In the case of the alpha particle, the energy
difference between the daughter states, the positive and negative
parity states, is reasonably large, although the intrinsic
single-particle states have large breaking of the parity and
charge symmetries.

\begin{figure}[h]
 %\centerline{\includegraphics[width=10cm]{he4density.EPS}}
 \centerline{\includegraphics[width=10cm]{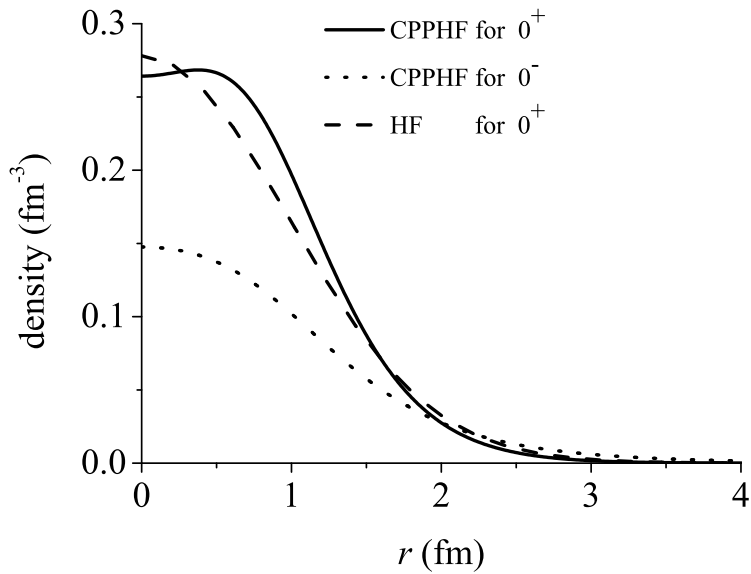}}
 \caption{Densities for the alpha particle with $x_{\rmT}=1.5$ and
 $x_{\rmTE}=0.81$ as a function of the radial coordinate $r$ (fm).
 The density of the $0^+$ state (the solid curve) and that
 of the $0^-$ state (the dotted curve) together with that of the
 $0^+$ state with the original Volkov No.~1 force in the simple
 Hartree-Fock scheme (the dashed curve) are shown.}
 \label{fig: density}
\end{figure}
Finally, we show the density distributions of the $0^+$ and $0^-$
states in Fig.~\ref{fig: density}. The density of
the $0^+$ in the CPPHF scheme decreases around the center and
increase in the middle region (around $r=0.8$ fm) from that of the
HF scheme. This change is induced by the admixture of the
$p$-state component. In fact, the wave function squared for the
$p$-state in Fig.~\ref{fig: wf} have the maximum around $r=0.8$
fm. Due to the smaller binding, the density of the $0^-$ state is
much more spread over the space than the case of the $0^+$ state.

\section{Summary}  \label{sec: summary}
We have developed a mean-field framework with the projection
method in order to treat the tensor force. To do this we mix
parities and charges into a single-particle state, instead of
using a single-particle state with a good parity and a definite
charge number. We perform the projections of parity and charge
from the intrinsic wave function, consisting of the parity- and
charge-mixed single-particle states. We take the variation of the
energy expectation value of the Hamiltonian with the projected
total wave function, and obtain a Hartree-Fock-like equation, the
charge- and parity-projected Hartree-Fock (CPPHF) equation.

We have applied the CPPHF equation to the alpha particle and shown
that the CPPHF method is able to treat the strong tensor
correlation. The variation after projection is very important to
obtain the tensor correlation in the present framework, because
the mixing of parity and charge does not occur if the parity and
charge projections are not performed before variation.

We have obtained a large tensor correlation energy in the CPPHF
method. The tensor correlation energy is about 31 MeV for
$x_\rmT=1.5$. To obtain such large tensor correlation energy, the
charge mixing and projection on top of the parity mixing and
projection is very important due to the isovector character of the
one-pion-exchange interaction. We have found that if we perform
the mixing of both charge and parity and their projections, the
tensor correlation energy becomes about three times larger than in
the case where only the parity mixing and projection is performed.
In the result of the alpha particle, we see that the $p$-state
component has the narrower width than the $s$-state component, and
does not seem like a simple $0p$-state. It means that to obtain
the correlation of the tensor force we need to mix higher-momentum
components. This fact indicates the increase of the kinetic energy
by the inclusion of the tensor correlation. In spite of the
increase of the kinetic energy, the correlation energy from the
tensor force produces more attractive energy and a net energy gain
is achieved in total. The probability of the $p$-state component
in the single-particle states becomes 10.5\% to 21.4\% with the
parameters adopted here. The probability of the total spin $S=2$
component, which corresponds to the $D$-state probability, changes
2.9\% to 13.3\% with the same parameters. As for the negative
parity state ($0^{-}$), the total binding energy becomes around
2.8 MeV. We have found that the density for the $0^+$ state
decreases around the center and increases in the middle-range
region due to the mixing of the $p$-state component. The density
distribution for the $0^-$ state is much broader than the ground
state ($0^+$).

In this paper, we have pursued a mean-field (single-particle)
framework with the projection method, in which the tensor force
can be treated in the same way as the central and LS forces. We
have made such a framework by combining the mixings of charge and
parity in a single-particle state and the charge and parity
projections on the total wave function. In the present study, we
have applied the CPPHF method only to the most simple closed-shell
nucleus, the alpha particle. The application of the CPPHF scheme
to heavier closed-shell nuclei, for example, $^{12}$C and
$^{16}$O, will show the role of the tensor force in the formation
of the shell structure of nuclei. The extension of the present
framework to the deformed case is also interesting and will give
us the insight of the role of the tensor force on the nuclear
deformation and the nuclear clustering. An application of the
present method on the relativistic mean field framework is under
progress \cite{ogawa04}.

\section*{Acknowledgments}
We acknowledge fruitful discussions with Prof. H.~Horiuchi and
Prof. Y.~Akaishi on the role of the pion in light nuclei. One of
the authors (SS) acknowledges the valuable suggestions of Prof.
M.~Kamimura on the numerical method. We are grateful to Prof.
I.~Tanihata for encouragements and interesting discussions on the
experimental consequences of the tensor correlation in finite
nuclei. The part of the present study is partially supported by a
Grant-in-Aid from the Japan Society for the Promotion of Science
(14340076). This work was partially performed in the Research
Project for Study of Unstable Nuclei from Nuclear Cluster Aspects
sponsored by the Institute of Physical and Chemical Research
(RIKEN).


\begin{thebibliography}{999}
\bibitem{vautherin72}
D.~Vautherin and D.M.~Brink, Phys. Rev.  C {5} (1972) 626.

\bibitem{decharge80}
J.~Decharg\'{e} and D.~Gogny, Phys. Rev. C {21} (1980) 1568.

\bibitem{walecka74}
J.D.~Walecka, Ann. of Phys. {83} (1974) 491; B.D.~Serot and J.D.~Walecka, in
{Advances in Nuclear Physics}, edited by J.W.~Negele and
E.~Vogt (Plenum Press, New York, 1986), vol.~16, p.~1.
\bibitem{sugahara94}
Y.~Sugahara and H.~Toki, Nucl. Phys. A{579} (1994) 557.

\bibitem{mayer49}
M.G. Mayer, Phys. Rev. {75} (1949) 1969;
O. Haxel, J.H.D. Jensen, and H.E. Suess, Phys. Rev. {75} (1949) 1766.

\bibitem{akaishi72}
Y.~Akaishi and S.~Nagata, Prog. Theor. Phys. {48} (1972) 133.

\bibitem{akaishi86}
Y.~Akaishi, in {Cluster Models and Other Topics}, edited by
T.T.S.~Kuo and E.~Osnes (World Scientific, Singapore, 1986),
p.~259.

\bibitem{carlson98}
J.~Carlson and
R.~Schiavilla,
Rev.~Mod.~Phys. {70} (1998) 743.

\bibitem{pieper01}
R.B.~Wiringa, S.C.~Pieper, J.~Carlson, and V.R.~Pandharipande,
Phys. Rev {C 62} (2000) 014001; S.C.~Pieper and R.B.~Wiringa,
{Annu.~Rev.~Nucl.~Part.~Sci.} {51} (2001) 53.

\bibitem{suzuki98}
Y.~Suzuki
and
K.~Varga,
{Stochastic Variational Approach to Quantum-Mechanical
Few-Body Problems}
(Springer-Verlag, Heidelberg, 1998), chap.~11.

\bibitem{terasawa60}
S.~Takagi, W.~Watari, and M.~Yasuno, Prog. Theor. Phys. {22} (1959) 549;
T.~Terasawa, Prog. Theor. Phys. {23} (1960) 87;
A.~Arima and T.~Terasawa,
Prog. Theor. Phys. {23} (1960) 115.

\bibitem{bethe71}
H.A.~Bethe, Annu. Rev. Nucl. Sci. {21} (1971) 93.

\bibitem{yukawa35}
H.~Yukawa,
Proc.~Phys.-Math.~Soc.~Jpn. {17} (1935) 48.

\bibitem{wong68}
C.W.~Wong, Nucl. Phys. {A108} (1968) 481.
\bibitem{tarbutton68}
R.M.~Tarbutton and K.T.R.~Davies, Nucl Phys. {A120} (1968) 1.

\bibitem{rouben72}
B.~Rouben and C.~Saunier, Phys. Rev. C {5} (1972) 1223.
\bibitem{brockmann78}
R.~Brockmann, Phys. Rev. C {18} (1978) 1510.
\bibitem{bouyssy87}
A.~Bouyssy, J.-F.~Mathiot, N.~Van~Giai, and S.~Marcos, Phys. Rev.
C {36} (1987) 380.

\bibitem{toki02}
H.~Toki, S.~Sugimoto, and K.~Ikeda, Prog. Theor. Phys. {108}
    (2002) 903.
\bibitem{amiet63}
J.-P.~Amiet and P.~Huguenin, Nucl. Phys. {46} (1963) 171.
\bibitem{amiet66}
J.-P.~Amiet and P.~Huguenin, Nucl. Phys. {80} (1966) 353.
\bibitem{rohl66}
W.H.~R\"{o}hl, Z. Phys. {195} (1966) 389.
\bibitem{bleuler66}
K. Bleuler, Proceedings of the International School of Physics
``Enrico Fermi'' Course 36, edited by C. Bloch (Academic Press,
Varenna, 1966), p. 464.
\bibitem{bassichis67}
W.H.~Bassichis and J.P.~Svenne, Phys. Rev. Lett. {18} (1967) 80.

\bibitem{takami96}
S.~Takami, K.~Yabana and K.~Ikeda, Prog Theor. Phys. {96} (1996)
407.

\bibitem{valor00}
A.~Valor, P.-H.~Heenen, and P.~Bonche, Nucl. Phys. {A671} (2000)
145.
\bibitem{rodriguez-guzman00}
R.~Rodr\'{i}guez-Guzm\'{a}n, J.L.~Egido, and L.M. Robledo, Phys.
Lett. {474B} (2000) 15.
\bibitem{rodriguez-guzman02}
R.~Rodr\'{i}guez-Guzm\'{a}n, J.L.~Egido, and L.M. Robledo, Nucl.
Phys. {A709} (2002) 201.
\bibitem{bender03}
M.~Bender and P.-H.~Heenen, Nucl. Phys. {A713} (2003) 390.

\bibitem{hiyama03}
E.~Hiyama, Y.~Kino, and M.~Kamimura, {Prog. Part. Nucl. Phys.}
{51} (2003) 223.

\bibitem{reinhard91} P.-G.~Reinhard, in {Computatinal Nuclear
Physics 1}, editted by K.~Langanke, J.A.~Maruhn, and S.E.~Koonin,
(Springer-Verlag, Heidelberg, 1991), p.~28; P.-G.~Reinhard and
R.Y.~Cusson, Nucl. Phys. A{378} (1982) 418.
\bibitem{volkov65}
A.~B.~Volkov, Nucl. Phys. {74} (1965) 33.

\bibitem{tamagaki68}
R.~Tamagaki, Prog. Theor. Phys. {39} (1968) 91.

\bibitem{oset82}
E. Oset, H. Toki and W. Weise, Phys. Rep. {83} (1982) 281.

\bibitem{kamada01}
H.~Kamada, A.~Nogga, W.~Gl\"{o}ckle, E.~Hiyama, M.~Kamimura, K.~Varga,
    Y.~Suzuki, M. Viviani, A.~Kievsky, S.~Rosati, J.~Carlson,
    S.C.~Pieper, R.B.~Wiringa, P.~Navr\'{a}til, B.R.~Barrett,
    N.~Barnea, W.~Leidemann, and G.~Orlandini, Phys. Rev. C {64}
    (2001) 044001.

\bibitem{vries87}
H.~de~Vries, C.W.~de~Jager, and C.~de~Vries, At. Data Nucl. Data
Tables {36} (1987) 495; I.~Sick, Phys. Lett. {116B} (1982) 212.

\bibitem{akaishi02}
Y.~Akaishi, private communication.

\bibitem{he4level}
D.R.~Tilley, H.R.~Weller, and G.M.~Hale, Nucl. Phys. A{541} (1992)
1.; Table of Isotopes, 8th edition, editted by R.B.~Firestone and
V.S.~Shirley (John Wiley \& Sons, New York, 1996).

\bibitem{ogawa04}
Y.~Ogawa, H.~Toki, S.~Tamenaga, H.~Shen, A.~Hosaka, S.~Sugimoto,
and K.~Ikeda, Prog. Theor. Phys. {111} (2004) 75.
\end{thebibliography}
\end{document}